\documentclass{aastex}
\begin{document}

\title{The multiphase nature of the intra--cluster medium of
some clusters of galaxies}

\author{Massimiliano~Bonamente$\,^{1}$, Richard~Lieu$\,^{2}$ and
Jonathan~P.~D.~Mittaz$\,^{3}$}

\affil{\(^{\scriptstyle 1} \)
{Osservatorio Astrofisico di Catania, Via S. Sofia 78, I-95125 Catania, Italy 
(mbo@sunct.ct.astro.it)}\\ 
\(^{\scriptstyle 2} \){Department of Physics, University of Alabama,
Huntsville, AL 35899, U.S.A. (lieur@cspar.uah.edu)}\\
\(^{\scriptstyle 3} \){Mullard Space Science Laboratory, UCL,
Holmbury St. Mary, Dorking, Surrey, RH5 6NT, U.K. (jpdm@mssl.ucl.ac.uk)}\\
}

\begin{abstract}
The discovery of EUV and soft X--ray excess emission 
in clusters of galaxies
(the cluster soft--excess phenomenon) 
challenged the notion of the hot ($\sim 10^{7-8}$ K) gas as the only dominant
thermal component of the intracluster medium (ICM). 
The spatial analysis of ROSAT PSPC 1/4 keV images 
presented here
reveals compelling evidence for substantial amounts of cold gas (HI) 
in the ICM of the Coma 
and Virgo clusters.
This finding bolsters  the original  interpretation of
the soft excess as emission from a `warm' ($\sim 10^6$ K) gas 
and points to the scenario of a multiphase ICM 
where the hot component co--exists with gases 
at sub--Virial temperatures (the `warm' and `cold' phases). 
\end{abstract}

\section{Introduction}

Studies of X-ray emission from the hot intracluster medium (ICM), using
the latest data and analysis methods, led many to infer that the
ICM may contain various gas components considerably cooler
than the Virial temperature (see, e.g., Allen 2000 and Buote et al. 
1999).  
The discovery of 
excess EUV and soft X-ray
emission from clusters of galaxies (Lieu et al. 1996a,b; 
Mittaz, Lieu, \& Lockman 1998) was originally
interpreted as due to substantial
amounts of warm gas at temperatures of $\sim 10^6$ K.
In this {\it paper} we present 
evidence for the widespread existence of even cooler gases, which
bolsters the possibility of a warm intermediate phase and the
notion of a multiphase ICM with sub-Virial temperature gases
playing an important role in a proper understanding of clusters.

Our results are obtained by a search for signatures of  spatially
resolved absorption in the
C-band images of the ROSAT PSPC (defined here and after as the passband 
between PI channels 20 and 41, or approximately 0.2 -- 0.4 keV).  
This band is ideally suited to the detection of
intervening cold ICM gas:
any depletion in the background cluster emission cannot be attributed to
such other possibilities as, e.g., warm absorbers.  
The C-band also
responds to small amounts of HI:
for incident radiation
from a plasma of 0.5 solar abundance and 
temperatures kT = 0.1 and 5.0 keV, reminiscent of the warm and hot ICM,
an optical depth in the C-band corresponds to cold gas of HI
column density N$_H \sim 1.3 \times 10^{20}$ and
$1.7 \times 10^{20}$ cm$^{-2}$, respectively.  Moreover, at high Galactic
latitudes the absorption of C-band extragalactic emission by our interstellar
medium  is not very severe, so that data of good statistical quality are
available for the brighter clusters.

\section{Method of analysis}
The method we
adopted is to evaluate the spatial smoothness of the soft excess
from two clusters, using PSPC C-band images.
The first step consists of estimating the hot gas
contribution to the C-band fluxes. To accomplish this, the cluster diffuse
emission was divided into concentric annuli and spectra in the R47 band
(a terminology often used to denote
PI channels 42-201, or energies 0.4 -- 2.0 keV) was modelled with a
photoelectrically absorbed thin plasma emissivity code (MEKAL in XSPEC),
where temperature (T) and elemental abundances (A) 
were fitted to the data (see Table 1).
In the R47 band, in fact, any contributions from a 
`warm' gas are minimal, and this
modelling returns a spatially resolved measurement of
the hot ICM parameters. At each radius $r$, the best-fit model  defines
a ratio $f(r)$ of the C-band to R47
count rates; removal of the hot ICM contribution 
from the C-band is obtained via the equation
\begin{equation}
c_{res}=c-f(r)\times h \, ,
\end{equation}
where $c$ and $h$ are respectively the background subtracted C-band and
R47
images.\footnote{Since azimuthal symmetry is
assumed, $f(r)$ is a radially symmetric `image' of
the C-to-R47 conversion factor described in the text.}
The distribution of the C-band residuals, $c_{res}$, 
reveals regions of soft excess
(positive values) as well as regions of absorption (negative values).
For the Coma cluster, a similar technique was adopted, whereby PSPC spectra
were succesfully modelled with a photoelectrically absorbed MEKAL code
with T and A fixed at the best-fit 
Ginga measurements (T=8.21 keV, A=0.21 solar,
Hughes et al. 1993). Given that the central region of the 
cluster  is sufficiently isothermal, the ratio f(r) is a constant
function of radius. 
 
After removal of the hot ICM contribution, 
a box of size 2.25$^{\prime}\times$ 2.25$^{\prime}$ scans the residual image area
($c_{res}$ in the notation of Eq. 1),
stepping its center by 0.25 arcmin at a time. The box size was chosen
to ensure sufficient enclosed counts
for normal distribution of Poisson fluctuations. Then,
if a background subtracted
image is smooth, its mean brightness should be zero and deviations
about the mean should follow a fixed gaussian.  This, a consequence of
the central limit theorem, was confirmed by our own simulations as well
as C-band images of blank fields.  As an example, we show in Figure 1
the results for various annuli of a blank
field within the central PSPC area; the
data were acquired during a 30 ksec pointing to a celestial direction where
the Galactic N$_{\rm H}$ is $\sim 1.4 \times 10^{21}$ cm$^{-2}$ 
(RA=50.29$^o$, DEC=40.74$^o$, RP number 800034a01). 
The C-band sky background is
low, due to its anti--correlation with HI, yet the forementioned
box size still yielded an average of $\sim$ 40 counts per box, so that
gaussian statistics apply.

When we apply the method to a cluster field, the X-ray emission
is often not smooth to begin with. Deviations from the azimuthally averaged 
surface brightness
for a R47 image of Coma are shown in Fig. 2a (see later for details 
on the observation):
the image clearly reveals substructures.  Nonetheless
our method of analysis remains meaningful, because such
regions of anisotropy
do not correspond to a different spectral hardness when compared
with the surroundings (i.e. the same conversion factor $f$ applies
when subtracting the hot ICM contribution to the C-band flux).  Further,
the lack of any resemblance between the soft excess and X-ray images,
as we demonstrate by Fig. 2b, confirms that any spatial structures in
the former are due to emission or absorption of the soft cluster flux.
For the Virgo cluster, similar results hold, with the exception of three
strong absorption features detected towards the cluster center, which are
positionally coincident with 
enhancements of the X-ray (0.4-2 keV) emission. These regions are the subject 
of a separate spectral analysis in the following section.

\section{The Virgo and Coma clusters}
The central region of the Virgo cluster is dominated 
by the
giant galaxy M87; near the center of Coma
are likewise
located the cluster's two brightest supergiant elliptical galaxies
(NGC 4874 and 
NGC 4889) and 
a few discrete X--ray sources.
We therefore excluded from analysis the innermost portions 
of the two clusters (angular radii $\leq$ 3 arcmin).
We show in Figure 3 the
C-band soft--excess image smoothness evaluation 
for the region between radii of 3 and
15 arcmin, centered at M87, and divided into 4 annuli. The dataset
is from the ROSAT archive (observation number RP8000365). 
Significant
departures from gaussian behavior, clearly noticeable within a radius of
9 arcmin, are in the form of an asymmetric extension to the left, which is
symptomatic of absorption features resolved by the scanning box\footnote{
Emission features, including any residual radial surface brightness
gradients
not removed by the subtraction of 
C-band radiation from the hot ICM, would have been manifested
as {\it right} extension tails, which are evidently absent.}.
Indeed the features within 6 arcmin radius are obviously identifiable
in the image.

A simple
`partial covering' model
to interpret the observations assumes that for each annulus
inwards of 12 arcmin,
a certain fraction of its area (which can be broken into smaller,
disjointed areas) has emission
uniformly shadowed by foreground
intrinsic HI, as illustrated in Fig. 4.  Then
the distribution of
soft emission is the sum of two gaussians: their peak positions
contain information about the 
mean brightness in the absence of intrinsic absorption and the HI
column for the absorbed areas.  Moreover, the relative normalization of the
gaussians converts to a partial covering factor.  
In Figure 5 we show the performance of this model for one
of the Virgo annuli, and in Table 2 
the best-fit parameters of the model are listed for
all the regions.  The decreasing
trend of the covering factor points to a
radially declining
influence of absorption {\it by clouds within the cluster}.
We emphasize that despite its success as
indicated by the $\Delta \chi^2$ values when compared
with  a single gaussian fit (Table 2), the model may still be
over-simplistic because the absolute $\chi^2$ values do not imply
acceptability.  Indeed,
additional spatially unresolved absorption may well
be present, and the effect of foreground cluster emission
may well have caused an underestimate of the HI towards these
central radii.  
It is also possible that the column density of HI clouds
is not uniform; nonetheless the quality 
of present data does not warrant more
sophisticated models, for this purpose one must await 
observations by the XMM/EPIC
instrument.

In Figure 6 
we show the three deepest absorption features which exist in the innermost
area; they are positionally coincident with prominent radio lobes 
(Harris et al. 1999) and with enhancements in the X--ray 
(0.5--2 keV) emission, see B\"{o}hringer et al. (1995), whose
study 
led to the proposition of a
lower temperature for the hot ICM in the radio lobes than
in the surrounding regions.
The forementioned enhancement was then interpreted as due to a 
Fe line feature at $\sim$ 1 keV.
We accordingly examined spectra for
two regions, one including one feature to the east and the other
to the south-west of M87. Since similar results apply to both regions,
here we focus on the
eastern knot: Figure 7a shows 
that a photo-absorbed MEKAL code with T=1.3 and A=0.45 
(as in B\"{o}hringer et al. 1995) can in fact model the observed spectrum 
($\chi^2$=188 for 155 degrees of freedom, with a null probability of 3.6 \%).
It is however not physically obvious why radio
features (symptomatic of  relativistic particles) are positionally coincident 
with a temperature decrease in the
hot gas.
An alternative explanation of the C-band  flux reduction
would invoke cold gas (HI): we therefore attempted to model the two regions
using the best-fit hot ICM parameters for their corresponding annulus
(T=1.47 keV, A=0.46 solar, see Table 1):
the eastern knot shows a -7$\sigma$ 
decrement from the model in the 0.2-0.4 keV band (fig. 7b),
yet if the total absorption was allowed to vary we could obtain a good
fit ($\chi^2$=165 for 154 degrees of freedom, null probability of 25 \%).
The extra absorption required above the Galactic line-of-sight column is
$N_H \sim$ 4.5$\times 10^{19}$ cm$^{-2}$ for a region
$\sim$ 2 square arcmin in size. 
A possible interpretation is that the high pressure of relativistic particles 
can lead to compression of the hot gas trapped inside the
radio lobes to high density, with consequent rapid 
cooling to very low temperatures.
The evidence for depleted emission  in other regions of the cluster
(Fig. 3), as well as
in the Coma cluster (see below), strengthens our 
present scenario, because
none of those are caused by the subtraction of spatially corresponding
enhancements in the hot ICM radiation.

A similar analysis of PSPC C--band data of the  Coma cluster 
(archival identification RP800005) proved equally fruitful.
The spatial distribution of the emission is clearly not smooth, see Fig. 8.
The use of a two--gaussian `partial covering' model
results in significant improvement of the goodness--of--fit
(see Table 3), where the decrease of the covering factor
with radius points  again to a centrally peaked distribution
of absorbers; the outermost annulus (12--15 arcmin)
in fact does not require a second gaussian.
As an example,
the performance of the two--gaussian model for the
6--9 arcmin region is shown in Fig. 9.
This result is of particular importance as Coma is the first
non--cooling flow cluster which exhibits evidence for cold gas clouds.

\section{Discussion and conclusions} 
The adopted method of analysis affords us estimates of
column densities and HI mass for
the
two clusters, see Table 4. 
Could this absorption be caused by line--of--sight
cluster galaxies? Rich clusters of galaxies, such as Virgo and Coma,
are known to have a large fraction of gas--poor
elliptical and S0 galaxies, and only a small fraction
of late type galaxies (such as spirals). Moreover, cluster
galaxies show an HI deficit when compared to 
field galaxies of same morphology, 
especially toward the central regions 
(e.g. Huchtmeier and Richter 1986; Dickey 1997).
In the case of Virgo a simple galaxy count 
for the 3--12 arcmin region from M87 (using the NED database)
revealed only three known, early type 
member galaxies along the line 
of sight, which is too few in number to account for
the absorption features 
of Fig. 6.  
Likewise, a recent 21-cm line investigation of the Coma cluster with the VLA array
(Bravo--Alfaro et al. 2000) detected no galaxies with HI content above $\sim 10^8$
M$_{\odot}$, the typical detection limit of that survey, in the central
10 arcminutes. 
Within the context of a galaxy origin of the cold gas,
this is again inconsistent with our HI estimates in Table 4.

It is far more plausible that the reported effect is due
to intracluster HI which
might have, at least in part, been released to the intergalactic space by
ram pressure stripping.      
Cold gas masses of the order 10$^{9-11} M_{\odot}$
contribute to only a small fraction 
of the cluster's total baryonic mass, 
the latter of order a few $\times 10^{12} M_{\odot}$ for Virgo (see, e.g.,
Bahcall and Sarazin 1979) and a few $\times 10^{14} M_{\odot}$ for Coma
(Briel, Henry and B\"{o}hringer 1992). 

We note that the phenomenon of the soft excess
radial trend (meaning rising importance of the soft component
with radius), known to exist in the clusters 
A1795 (Mittaz, Lieu and Lockman 1998)
and A2199 (Lieu et al. 1999), was recently interpreted as indicative
of centrally peaked intrinsic cluster absorption 
(Lieu, Bonamente and Mittaz 2000).  As shown in Figure 10, 
this effect is also present in the C--band excess of Virgo,
and may indeed be due to the gradual disappearance of
absorption with increasing radius reported in Table 4.

The co--existence  of a cold phase
and hot ICM components raises questions concerning 
{\it if} and
{\it where} an intermediate  warm phase is present.
The evidence for cold HI presented in this {\it paper} reinforces
the thermal interpretation of the cluster soft--excess syndrome:
some warm gas can certainly be generated at the interface between
the cold and hot gases, e.g. by the `mixing layer' mechanism 
(Fabian 1997).
The emerging scenario is a three--phase ICM,
where the hot gas emits the bulk of the
X--ray emission, and the warm gas is responsible
for the soft--excess emission.
The cooler gas (T $\leq 10^5$ K), found by our spatial smoothness
test, causes substantial reduction of EUV and soft X--ray fluxes 
in some locations, and
the soft excess radial trend detected in the C band of Virgo (see Fig. 10) could then
be due to the absorption effect of a centrally
peaked distribution of HI. 

\newpage
\section*{References}
\noindent
Allen 2000, {\it MNRAS accepted}. \\
\noindent
Bahcall, J.N. and Sarazin, C.L. 1977, {\it ApJ}, {\bf 213}, L99.\\
\noindent
B\"{o}hringer, H., Nulsen, P.E.J., Braun, R. and Fabian, A.C. 1995,\\
\indent {\it MNRAS}, {\bf 274}, L67.\\
\noindent
Bravo--Alfaro, H., Cayatte, V., Van Gorkom, J.H. and Balkowski C. 2000, \\
\indent {\it AJ}, {\bf 119}, 580. \\
\noindent
Briel, U.G., Henry, J.P. and B\"{o}hringer, H. 1992 {\it A\&A}, {\bf 259}, L31.\\ 
\noindent
Buote, D. A., Canizares, C. R. and Fabian, A. C. 1999, {\it MNRAS} , {\bf 310}, 483.\\
\noindent
Dickey, J.M. 1997, {\it AJ}, {\bf 113}, 1939. \\
\noindent
Fabian, A.C. 1997, {\it Science}, {\bf 113}, 48. \\
\noindent
Harris, D.E., Owen, F., Biretta, J.A. and Junor, W. 1999, 
{\it Proceedings of the \\
\indent Ringberg Workshop on Diffuse Thermal and Relativistic
Plasma in Galaxy Clusters},\\
\indent  MPE Report 271, 111. \\
\noindent
Huchtmeier, W.G. and Richter, O.G. 1986, {\it A\&A}, {\bf 64}, 111. \\
\noindent
Hughes, J.P., Butcher, J.A., Stewart, G.C. and Tanaka, Y. 1993, {\it ApJ}, \\
\indent {\bf 404}, 611.\\
\noindent
~Kaastra, J.S. 1992 in \it An X-Ray Spectral Code for Optically Thin Plasmas \rm \\\indent
(Internal SRON-Leiden Report, updated version 2.0) \\
\noindent
Lieu, R., Mittaz, J.P.D., Bowyer, S., Lockman, F.J., Hwang, C.-Y. and \\
\indent Schmitt, J.H.H.M. 1996a, {\it ApJ}, {\bf 458}, L5.\\
\noindent
 Lieu, R., Mittaz, J.P.D., Bowyer, S., Breen, J.O.,
Lockman, F.J., \\
\indent Murphy, E.M. \& Hwang, C. -Y. 1996b, {\it Science}, {\bf 274},1335--1338. \\
\noindent
Lieu, R., Bonamente, M., Mittaz, J.P.D., Durret, F., Dos Santos, S. and \\
\indent Kaastra, J. 1999, {\it ApJ}, {\bf 527}, L77.\\
\noindent
Lieu, R., Bonamente, M. and Mittaz, J.P.D. 2000, {\it A\&A submitted}. \\
\noindent
~Mewe, R., Gronenschild, E.H.B.M., and van den Oord, G.H.J., 1985 \\
\indent {\it A\&A Supp.}, {\bf 62}, 197--254.  \\
\noindent Mewe, R., Lemen, J.R., and van den Oord, G.H.J. 1986,
\it A\&A Supp.\rm, \\
\indent {\bf 65} 511--536. \\
\noindent
~Mittaz, J.P.D., Lieu, R., Lockman, F.J. 1998, {\it ApJ}, {\bf
498},
L17--20. \\
\noindent
 Morrison, R. and McCammon, D. 1983, {\it ApJ}, {\bf 270}, 119.\\

\newpage

\begin{table}[h!]
\begin{center}
\tighten
\caption{Best--fit parameters for the hot ICM of Virgo; errors are
90\% confidence ($\chi^2$ + 2.701 criterion).}
\vspace{22pt}
\small
\begin{tabular}{lccc} \hline
Region (arcmin) & Temperature (keV) & Abundance & red. $\chi^2$
(d.o.f) \\
 & & & \\
\hline
0--3 & 1.47 $\pm^{0.05}_{0.04}$ & 0.46 $\pm^{0.03}_{0.04}$ & 1.15(156) \\
3--5 & 2.03 $\pm^{0.12}_{0.13}$ & 0.55 $\pm^{0.09}_{0.08}$ & 1.25(156) \\
5--7 & 2.42 $\pm^{0.24}_{0.19}$ & 0.45 $\pm^{0.09}_{0.07}$ & 1.2(156) \\
7--10 & 2.53 $\pm^{0.21}_{0.22}$ & 0.39 $\pm^{0.07}_{0.08}$ & 1.1(156) \\
10--15 & 2.49 $\pm^{0.19}_{0.19}$ & 0.33 $\pm^{0.06}_{0.08}$ & 1.29(156) \\
15--19 & 3.14 $\pm^{0.42}_{0.34}$ & 0.37 $\pm^{0.11}_{0.10}$ & 1.07(156) \\
 & & & \\
\hline
\end{tabular}
\end{center}
\end{table}

\begin{table}[h!]
\begin{center}
\tighten
\caption{Performance of the two--gaussian model for the Virgo cluster, 
where $\mu_1$ and $\mu_2$
are the means of the two gaussians, each having fixed width at  $\sigma$=1; 
$\Delta \chi^2$
is the $\chi^2$ reduction from the one--gaussian fit, and 
the covering factor is obtained by the ratio of the 
normalization constants. Errors are 90 \% confidence ($\chi^2$+ 2.701
criterion). The outermost annulus 
does not require a second (left) gaussian.
\label{tab:virgo}}
\vspace{22pt}
\small
\begin{tabular}{cccccc}
\hline 
 Region (arcmin)  & $\mu_1$ & $\mu_2$ & cov. factor 
& $\chi^2$(d.o.f) & $\Delta \chi^2$ \\
\hline 
3--6 & 0.93$\pm 0.09$ & 4.1$\pm^{0.06}_{0.05}$ & 0.34$\pm 0.02$ & 67.6(26) & 594.4 \\
6--9 & 2.27$\pm^{0.15}_{0.13}$ & 3.27$\pm^{0.04}_{0.05}$ & 0.16$\pm 0.02$ &
  19.4(17) & 76.4 \\
9--12& 1.7$\pm^{0.4}_{0.25}$ & 3.07$\pm 0.03$ & 0.042$\pm 0.019$ &
  31.5(16) & 7.5 \\
12--15& 2.32$\pm 0.017$ & -- & -- & 42.8(18) & -- \\
\hline    
\end{tabular}
\end{center}
\end{table}

\begin{table}[h!]
\begin{center}
\tighten
\caption{Two--gaussian model for the Coma cluster,
see caption of Table 1.  
\label{tab:coma1}}
\vspace{22pt}
\small
\begin{tabular}{cccccc}
\hline 
 Region (arcmin)  & $\mu_1$ & $\mu_2$ & cov. factor
& $\chi^2$(d.o.f) & $\Delta \chi^2$ \\
\hline 
3--6  & 3.16$\pm 0.06$ & 5.3$\pm 0.084$ & 0.58$\pm 0.03$ & 15.7(16) & 302.8 \\
6--9  & 1.35$\pm 0.08$ & 3.5$\pm 0.045$ & 0.32$\pm 0.02$ & 50(21)   & 379.8 \\
9--12 & 0.14$\pm 0.15$ & 1.85$\pm 0.04$ & 0.14$\pm 0.027$& 10.4(17) & 85.4 \\
12--15& 1.05$\pm 0.08$ & --             & --             & 21(14)   & --   \\
\hline
\end{tabular}
\end{center}
\end{table}

\newpage 

\begin{table}[h!]
\begin{center}
\tighten
\caption{HI column density ($N_{\rm H}$) and mass estimates for
the Coma and Virgo clusters.
The peak positions of the two gaussians (Tables 1 and 2) are
first transformed to units of  flux, and then, using the absorption cross--section
of Morrison and McCammon (1983), the flux difference is converted to
$N_{\rm H}$.
The tabulated mass refers to the total value for the lines of sight 
within each annular region as $ M=N_{\rm H}\times (area) \times (c. f.) \times
m_{\rm H}$, where $c. f.$ is 
the covering factor of the region as from Tables 1 and 2.
\label{tab:coma2}}
\vspace{22pt}
\small
\begin{tabular}{cccc}
\hline 
Cluster & region (arcmin) & $N_{\rm H}$ (cm$^{-2}$) & mass ($M_{\odot}$) \\
\hline 
Virgo   & 3--6  & 3.5 $\times 10^{19}$ & 3.4 $\times 10^8$ \\
 "      & 6--9  & 3 $\times 10^{19}$ & 2.4 $\times 10^8$ \\
 "      & 9--12 & 6 $\times 10^{19}$ & 1.7 $\times 10^8$ \\
\hline 
Coma   & 3--6 & 1.4 $\times 10^{19}$ & 3.8 $\times 10^9$ \\
"      & 6--9 & 1.5 $\times 10^{19}$ & 4 $\times 10^9$ \\ 
"      & 9--12 & 1.7 $\times 10^{19}$ & 1.8 $\times 10^9$ \\
\hline
\end{tabular}
\end{center}
\end{table}

\newpage 

Figure 1: 
Spatial distribution of events for
the inner annuli of a blank field centered at PSPC boresight.
A 2.25$^\prime \times 2.25^{\prime}$ scanning box detects, 
for each detector position,
the deviation z between the number of enclosed counts (N) 
and the mean background level ($\mu$), 
in units of the standard deviation $\sigma$
(where z=(N-$\mu$)/$\sigma$).
The y--axis gives total number of boxes at
a given deviation z and
the dashed line is the best--fit profile expected from a smooth
distribution, viz., a gaussian of fixed width $\sigma$=1. The gaussian
mean was varied to account for any possible systematics
in the background subtraction.

Fig. 2: 
(a) At each detector position, a 2.25$^{\prime} \times 2.25^{\prime}$
scanning box detects the difference between the number of enclosed 
R47 counts and the
azimuthally averaged value at that radius, in units 
of the standard deviation (see caption of Fig. 1). 
(b) C-band residual image of Coma: same scanning box detects, 
at each detector position, the deviation between
the number of enclosed counts and 0, the mean expected value in the absence
of absorption or soft excess emission. Unit of
measure is the standard deviation $\sigma$.  The $\sigma$-scale
of 2(b) is set to reveal the absorption features responsible
for the left tail of Figure 3;
these features are clearly
not positionally coincident with any emission features of the
hot ICM, as is evident from a comparison of 2(a) with 2(b).
Regions containing obvious point sources were excluded from the analysis (gray boxes).
 
Figure 3:
Spatial distribution of events for the central region of the
 Virgo cluster. As in Fig. 1, dashed line gives the best--fit
model for a smooth distribution
(i.e. one--gaussian) except that it fails to account for the
data here. Annular radii are measured from the position of M87, which is at
PSPC boresight.

Figure 4: The scanning box (light grey) can detect silhouettes
(dark grey) of the surface brightness which are responsible for the 
bimodal behavior of the spatial distribution of events (see e.g. Fig. 4 and
Fig. 7).

Figure 5:
Fitting the distribution of events in the 6--9 arcmin annulus of Virgo with
the two--gaussian `partial covering' model. A significant improvement
over one gaussian is achieved (see also Table 1). The positive mean of the
right (main)  gaussian revelas soft--excess emission.

Figure 6:
C--band soft excess image  of Virgo,
in units of statistical significance $\sigma$ of signals within 
 a 2.25$^\prime \times 2.25^{\prime}$ 
  scanning box centered at each pixel position.
The deep
absorption features to the south of M87 (cross) are coincident with
the location of radio lobes (Harris et al. 1999). 

Figure 7: 
(a) PSPC spectrum of a $\sim$ 2 square arcmin region encompassing the
`absorption feature' to the east of M87 (cross), fitted to a MEKAL code with
the parameters of B\"{o}hringer et al. (1995).
(b) same PSPC spectrum when only PI channels 42-201 ($\sim 0.4-2$ keV)
are fitted to a MEKAL code with the best-fit parameters of the
0-3 arcmin annulus (see Table 1).

Figure 8:
Event distribution for the Coma cluster (see caption of Fig. 3); the
dashed line is the best--fit single gaussian model, as before. 
The center of our annular system is at the X--ray centroid
of Coma, which is near boresight. The positive mean of the right
gaussian reveals soft excess emission.

Figure 9:
Use of the `partial covering' model for the 6--9 annulus 
of Coma brings considerable improvement to the fit (see also Table 2). 

Figure 10:
The `soft-excess radial trend'
 effect of the Virgo cluster, illustrated by a plot
against cluster radius of the soft X-ray fractional excess $\eta$,
defined as $\eta = (p - q)/q$, where for a given
annulus $p$ is the observed
C--band
flux
after subtracting the sky background, and $q$ is the
expected flux from the hot ICM as determined by fitting the
PI channels 50 -- 200 ($\sim$ 0.5 -- 2.0 keV)
using the MEKAL thin plasma emission code
(Mewe, Lemen and van den Oord 1986; Mewe, Gronenshild and van den Oord 1985;
Kaastra 1992)
and Galactic absorption according to Morrison \& McCammon (1983).

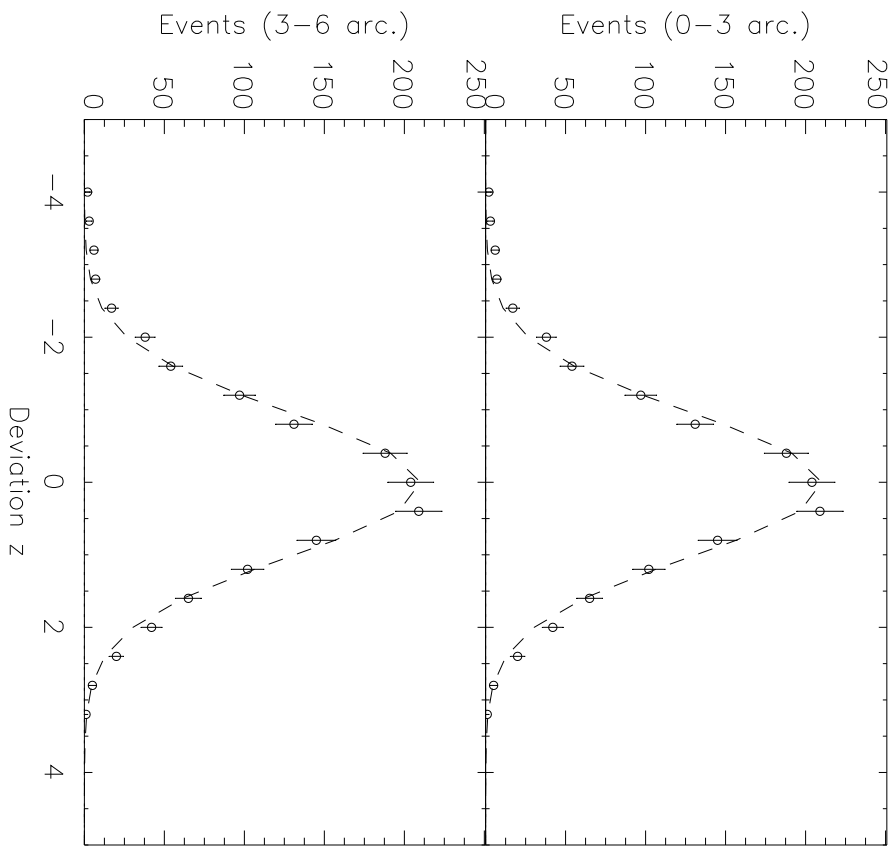
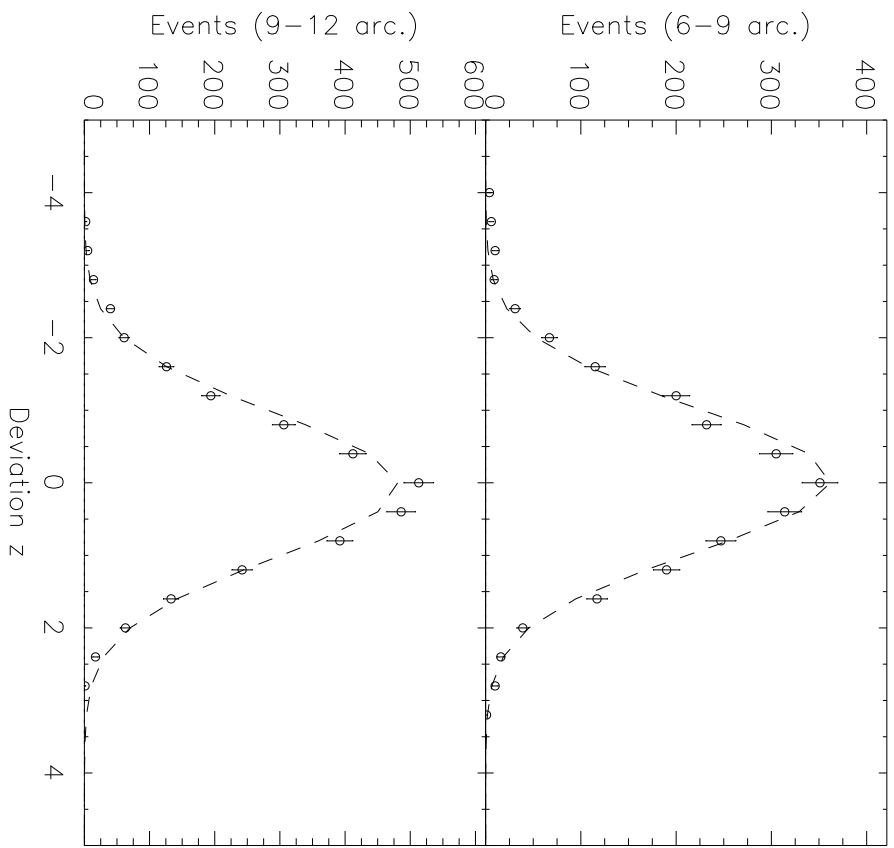
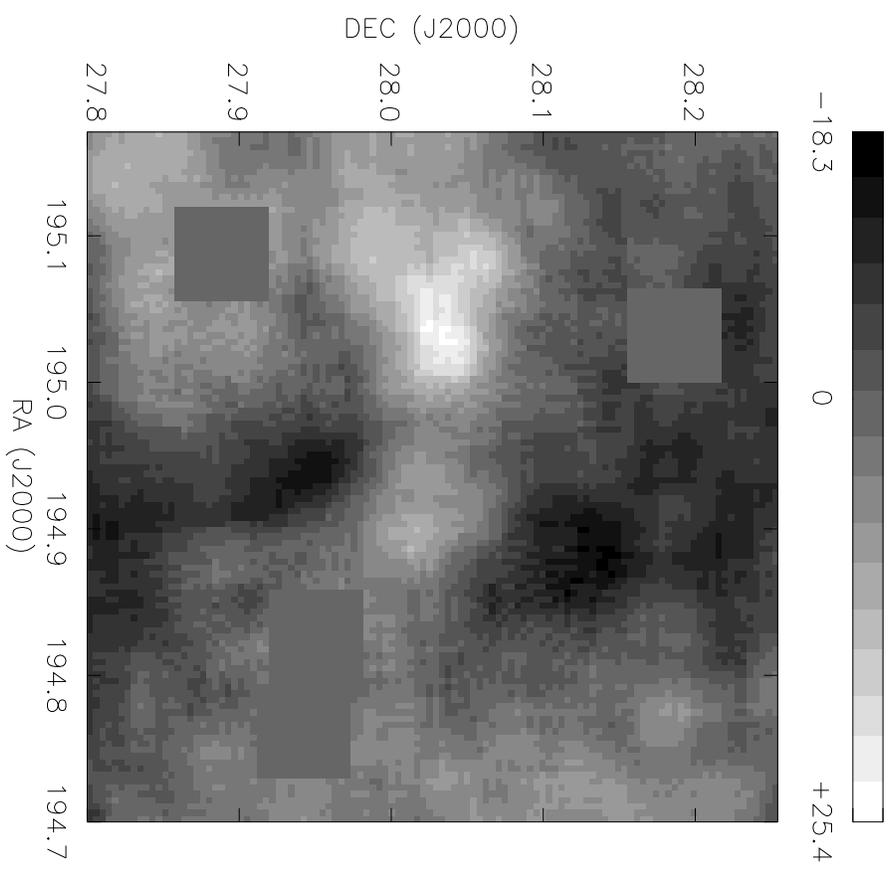
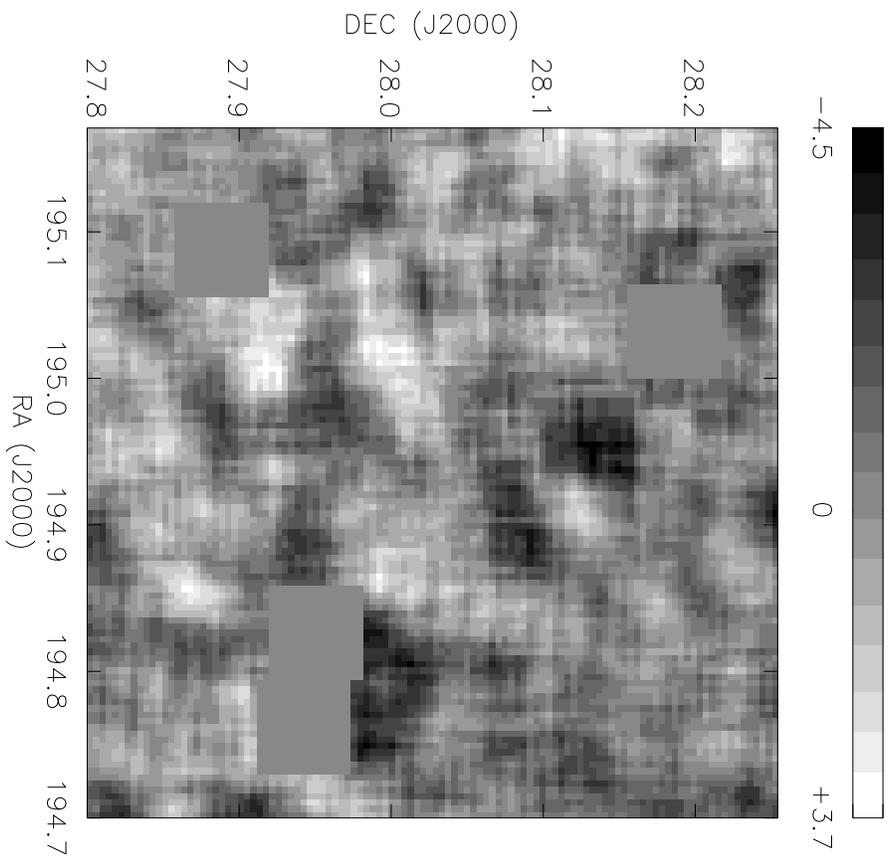
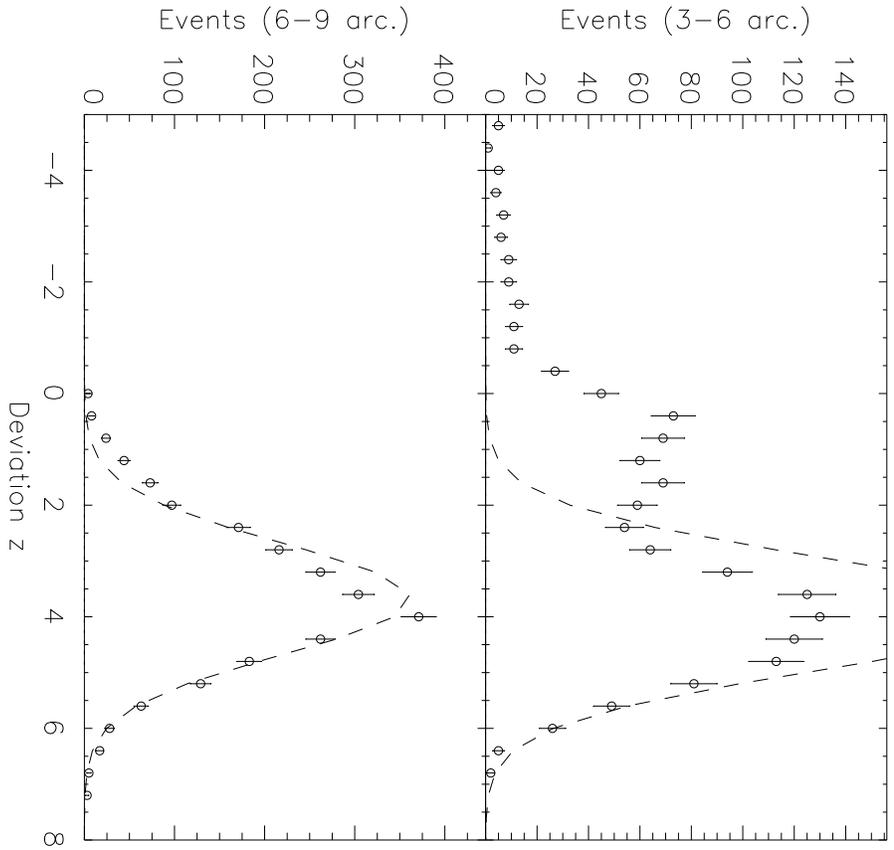
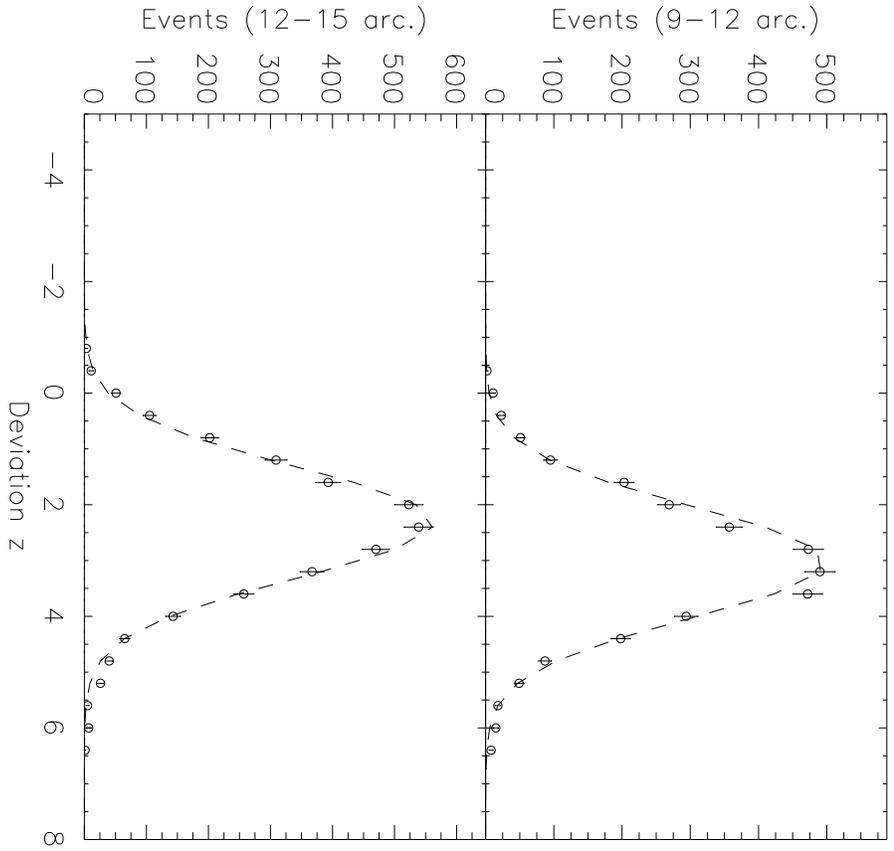
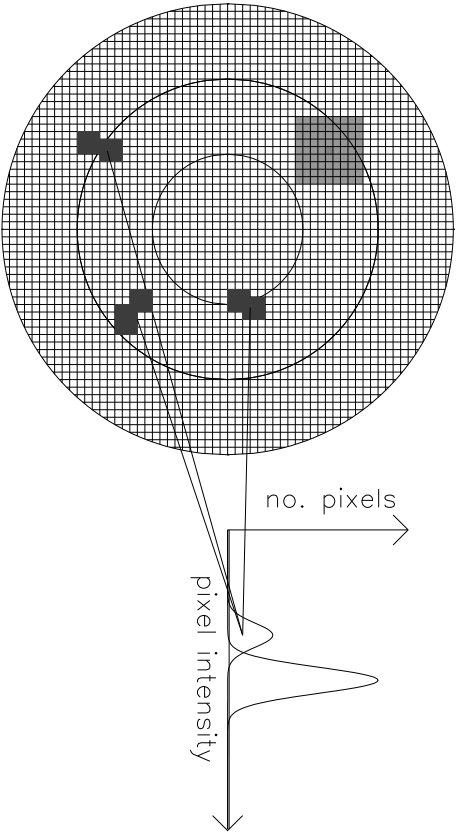
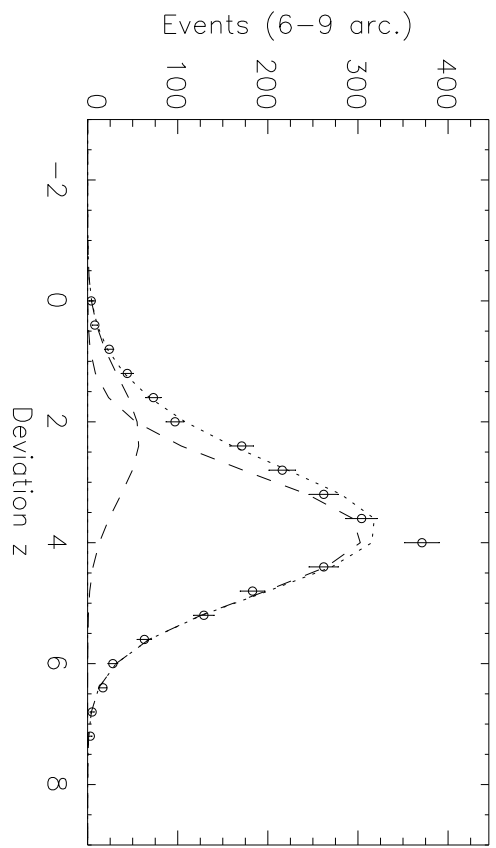
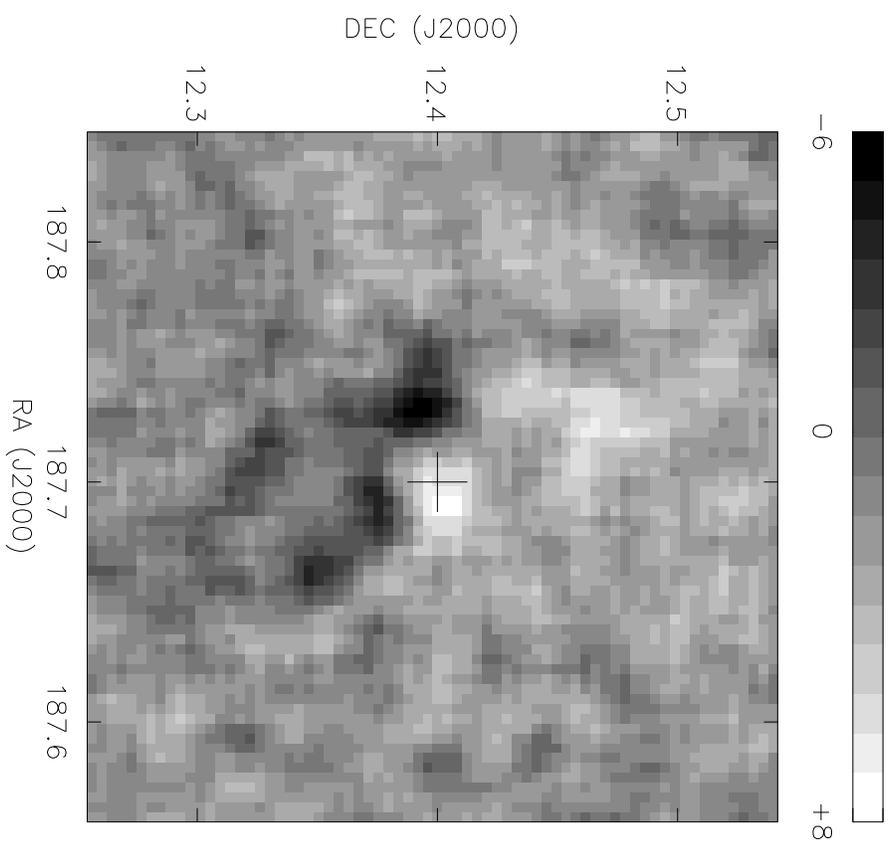
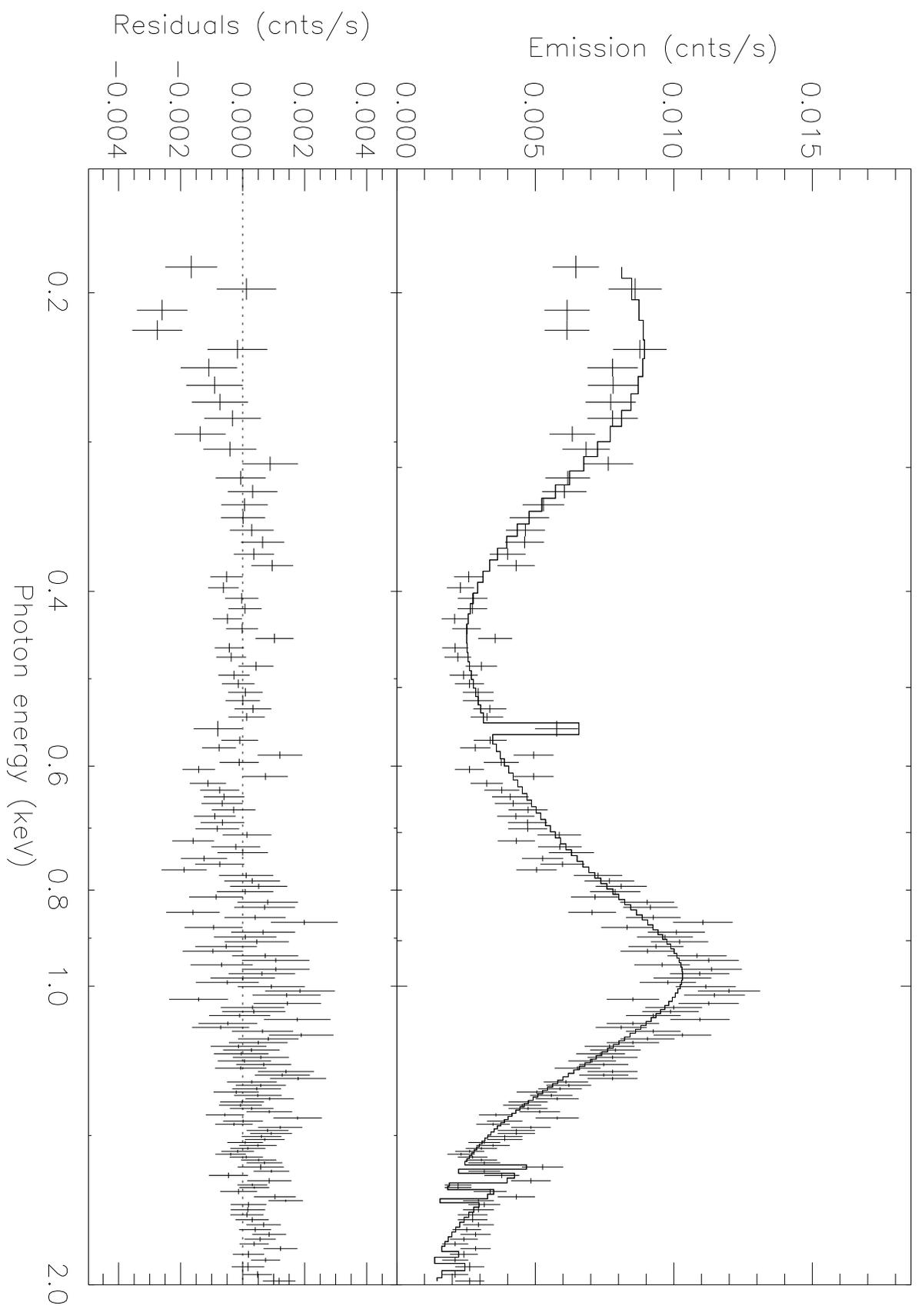
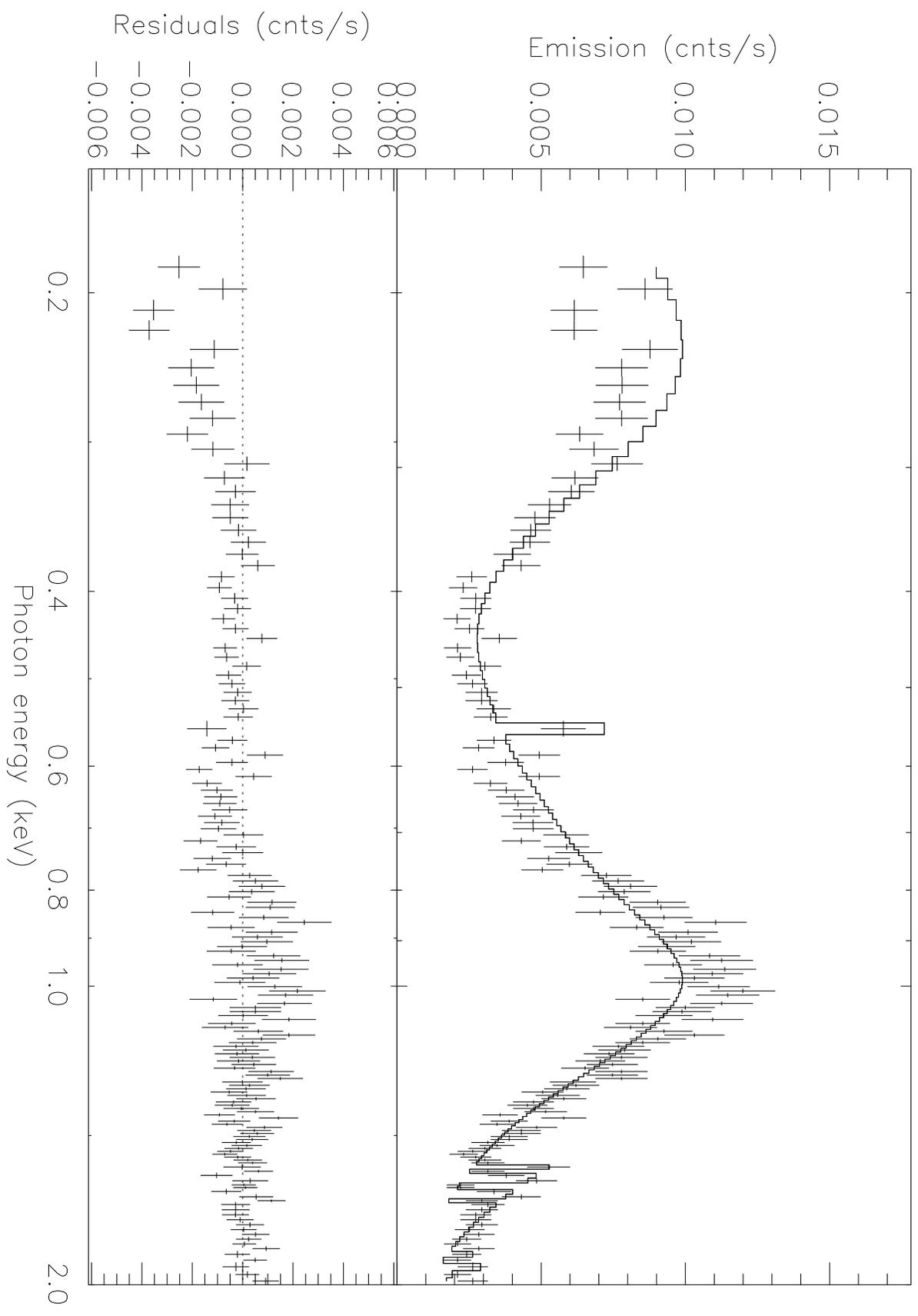
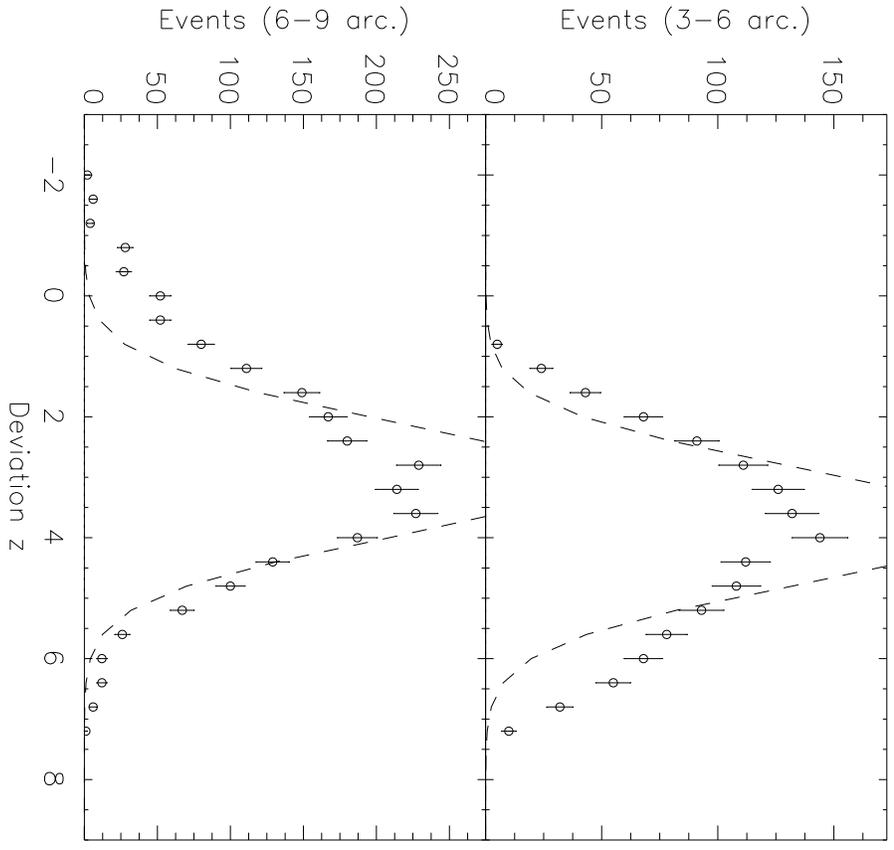
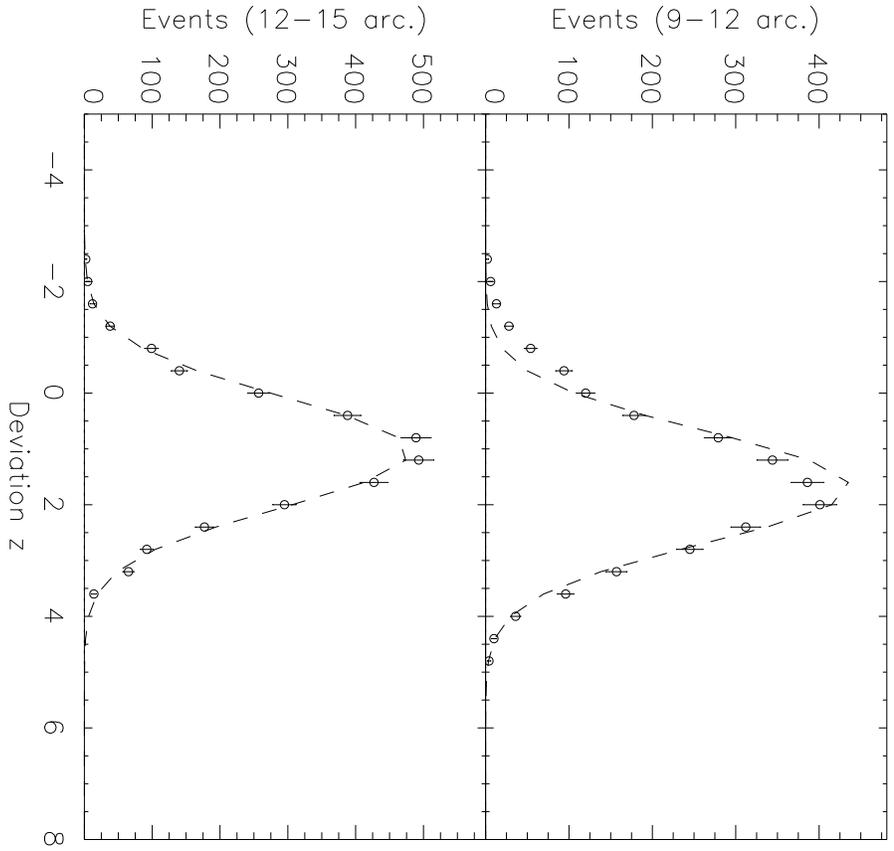
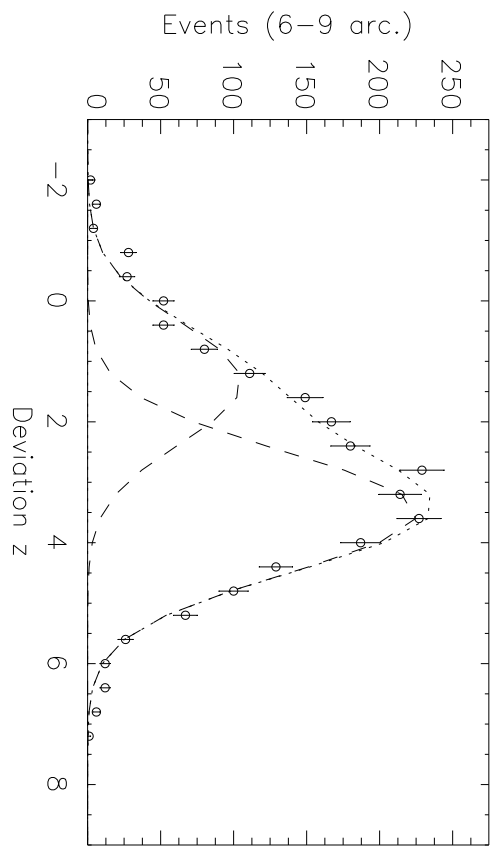
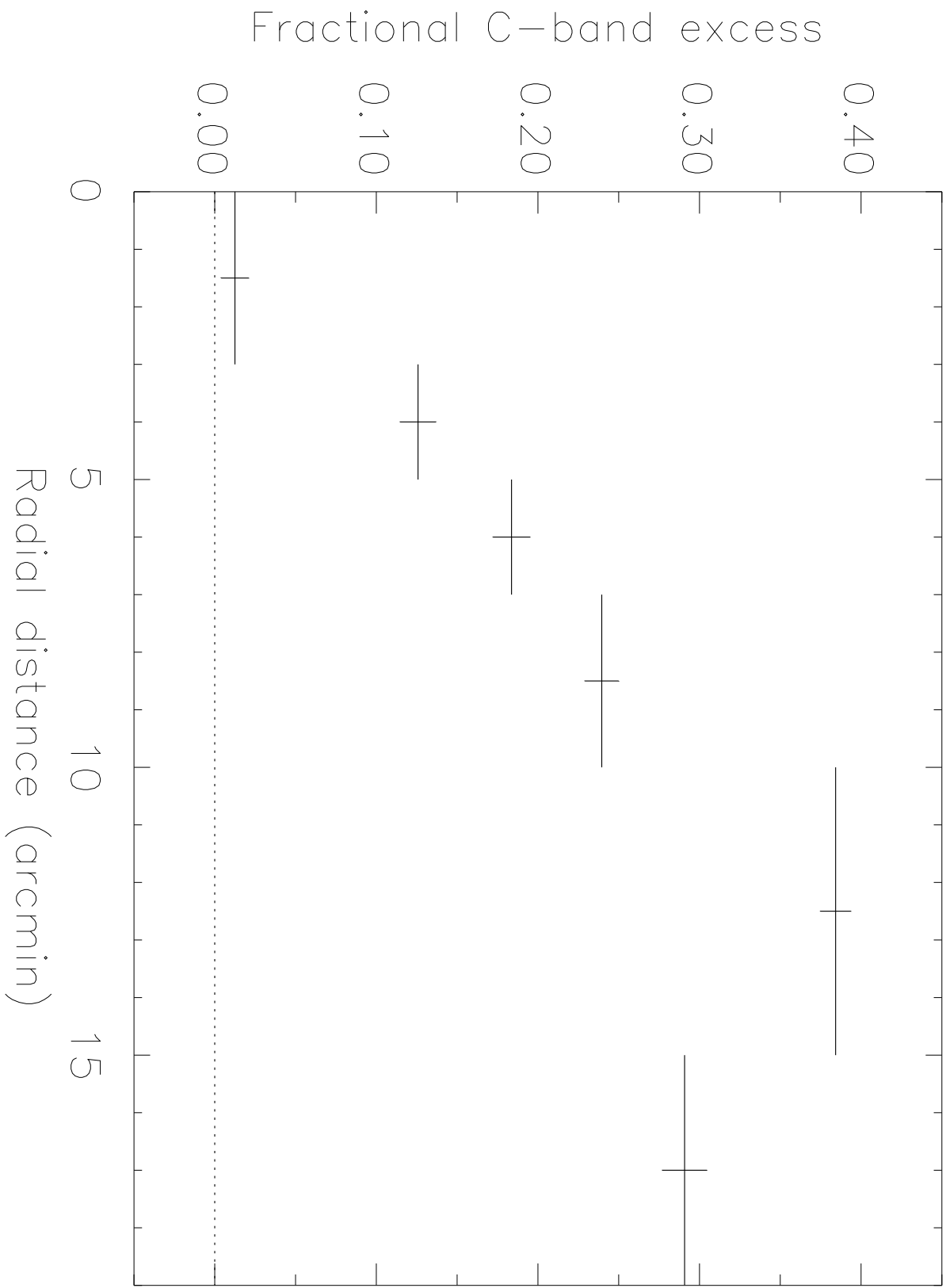
\end{document}